# Towards High-Resolution Orbitrap Mass Spectrometry for Next-Generation In-Situ Space Dust Analysis

M. Malečková,[1] I. Zymak,[1] A. Spesyvyi,[1,2] M. Polášek,[1] B. Cherville,[1] M. Lacko,[1] M. Ferus,[1] A. Charvat,[2] J. Žabka,[1] and B. Abel[1,2*]

1) J. Heyrovský Institute of Physical Chemistry of the Czech Academy of Sciences, Prague, 18223, Czech Republic
2) Institute of Chemical Technology, Leipzig University, Leipzig, 04103, Germany

## Abstract

Cosmic and planetary dust hold vital clues to the chemical evolution of the solar system, yet *in situ* analysis of their molecular and elemental composition remains technically challenging. Here we present laboratory results from HANKA (*High-resolution mass Analyzer for Nano-scale Kinetic Astro materials*), a compact Orbitrap™-based mass spectrometer developed towards the goal of a universal dust detector for space applications. Using infrared laser ablation and plasma formation under vacuum, we analyzed solid-state samples representative of Lunar, Martian, and Meteoritic material. The resulting mass spectra—recorded at resolving powers of ~60,000 (FWHM)—reveal complex, but characteristic element mixtures. These results demonstrate HANKA's single event sensitivity, and its mass resolution with the ability to resolve complex mass spectra and possibly differentiate geochemical signatures across planetary bodies. The system's dynamic range of 3-4 orders of magnitude enables even the detection of trace compounds. The minimalistic impact sampling approach enables fast, high-precision compositional mass analysis without complex sample preparation, making the instrument well-suited for orbital, surface, or flyby missions with expected dust impacts. Our system consists of an IR-laser system that simulates impacts of nano to micron sized solid-state dust and ice particles from space. It therefore allows to perform so called analogue experiments in the laboratory. We conclude that IR- laser pulses on a solid-state material are not only well suited for ice particle impact analogues but also a good analogue experiment for solid state particle impacts on a dust detector electrode in space. Moreover, the IR laser is employed only for an optimal analogue experiment simulating the impact event - it is not necessary in space on a spacecraft detecting space dust. This puts our approach apart from laser ablation mass spectrometry applications employing a UV laser in space (e.g., on rovers) and provides a unique selling point here.

___________
*Email: bernd.abel@uni-leipzig.de, bernd.abel@jh-inst.cas.cz

## 1. Introduction

In the vast expanse of space, micro- and nanoscopic dust particles on the one hand serve as messengers from the early solar system, carrying invaluable information about the formation and evolution of planetary bodies. These particles, originating from comets, asteroids, and interstellar space,[1] are rich in both organic and inorganic compounds. Analyzing their composition provides insights into planetary formation processes, the delivery of prebiotic materials to Earth, and the potential for life elsewhere in the universe.[2] On the other hand, dust particles are observed near planets of our solar system or their moons. In the latter case, they contain a significant amount of ice and they are emitted from ice moons, such as Enceladus.[2]

Traditional dust analyzers deployed in space missions, such as time-of-flight (TOF) mass spectrometers, have been instrumental in advancing our understanding of cosmic dust of different origin. This led to the rapidly expanding science field of dust astronomy.[2, 3] Instruments like the Cosmic Dust Analyzer[4] (CDA) aboard the Cassini spacecraft[5] and the Cometary Secondary Ion Mass Analyzer (COSIMA)[6] on the Rosetta mission have provided valuable data on the composition of dust particles in the Saturnian system and comet 67P/Churyumov-Gerasimenko, respectively.[7-11] A new generation of instruments such as SUDA (*SUrface Dust Analyser*)[12] is on board of Europa Clipper that started to the ice moons of Jupiter in 2024. Powerful TOF mass analyzers have also been proposed for other missions.[13, 14] However, these instruments often face limitations in mass resolution and sensitivity, hindering the unambiguous identification of complex organic molecules and isotopic variants.

The advent of Orbitrap mass spectrometry has revolutionized analytical capabilities in laboratory settings,[15] offering ultra-high mass resolution (up to 1,000,000 FWHM at $m/z$ = 200) and sub-ppm mass accuracy.[16, 17] These features enable the precise differentiation of isobaric species and the detection of trace compounds in complex mixtures. Recognizing the potential of Orbitrap technology for space applications,[18-21] recent efforts have focused on miniaturizing and adapting these instruments for extraterrestrial environments.[19, 20, 22-25]

Recent developments in Orbitrap-based mass spectrometry have opened the door to a new era of ultra-high-resolution analysis in space.[19, 20, 26, 27] Originally developed for laboratory[28] and proteomics applications,[29, 30] the Orbitrap™ analyzer provides mass resolutions exceeding 100,000 (FWHM) and sub-ppm mass accuracy, enabling the precise identification of even trace-level and structurally similar compounds.[31] Now, with efforts to miniaturize and ruggedize this technology, a new generation of compact Orbitrap instruments[16] is emerging—designed to operate in the harsh and sample-limited conditions of space.

In this context, we introduce *HANKA*—the *High-resolution mass Analyzer for Nano-scale Kinetic Astro materials*—our newly developed compact Orbitrap-based instrument designed specifically to analyze cosmic dust with very high mass resolution.[32] HANKA aims to become a *universal dust detector for space applications*, capable of capturing and deciphering the full chemical fingerprint of individual dust particles and their instrument electrode impacts, either from planetary surfaces/interiors, cometary tails, or the interstellar medium.

As a proof of concept, we conducted a series of laboratory experiments using infrared laser absorption/ablation and plasma ionization (ILAPI) of 3 types of meteoritic samples, including lunar and Martian materials, as well as a water target under vacuum conditions – with the latter simulating impacts and cold plasma ionization of ice particles. For the experiments, we selected three meteorite samples: the Martian shergottite NWA 12269, the lunar melt breccia NWA 11444, and the ordinary chondrite of the L3 group RaS 445. Basic information on these samples is available in the Meteoritical Bulletin Database (Refs. 34-36). Impacts of ice particles containing organic molecules (such as simple amino acids) were simulated and analyzed by ILAPI of liquid water samples containing organic molecules, our standard analogue experiments for ice particle impacts in the past. The clear advantage of IR laser radiation (over UV ablation and ionization) and the ILAPI method – as will be shown here for the first time – is its ability to simulate both, impact and cold plasma ionization of solid-state mineral particles and ice particles, with the same method. This strategy is now employed to generally simulate hypervelocity dust impacts of various nature, and to sample the generated ions and to perform a subsequent highly accurate and sensitive mass analysis. The mass analyzer with a dedicated ion optics is now an Orbitrap-type of mass detector. With this paper we want to demonstrate HANKA's ability to resolve the resulting impact ionized material (via ILAPI) with a mass resolution of ~60,000 (FWHM). The measured spectra reveal complex but characteristic element compositions and validate both the sensitivity and resolution of HANKA's mass detection capabilities. This high-resolution performance, combined with a certain robustness and size (scalability), may position HANKA as a next-generation tool for future missions. Its compact design and impact sampling approach (without any laser in space applications) make it ideal for applications with orbiters, flyby probes, and even CubeSats—enabling the real-time, in situ analysis of dust across the solar system, without the need for sample return or complex sample preparation. With an IR laser, on the other hand, it may be employed also for rover missions.

In this study, we present the scientific motivation, preliminary experimental validation, and HANKA's potential for laboratory analogue experiments and even future spaceflights. Its advanced version may even become a universal platform for dust astronomy, space mining, as well as exploring biosignatures. By extending laboratory performance towards mission-ready engineering in the future (in progress at present), HANKA may have the potential to redefine what's possible in the mass analysis of nano- to micron scale cosmic materials in the future.

## 2. Materials and Methods

### 2.1 HANKA mass analyzer

The HANKA mass analyzer was designed to fit the CubeSat platforms due to high flexibility of such satellite sub-systems. The laboratory prototype was implemented using *cubic modular vacuum chambers* (6x6x6 inches, Ideal Vacuum Products LLC, USA) to ensure dimensional compatibility. The instrument setup, as shown in Figure 1, comprises a laser ablation/ionization chamber interfaced to the mass analyzer through switchable guiding electrodes and transfer ion optics. The mass analyzer consists of a differentially pumped cell with *Orbitrap* electrodes assembly (D30, Thermo Fisher Scientific Inc., US) connected via a preamplifier (JanasCard, *Czechia*) to an analog-to-digital (ADC) acquisition device converter (PicoScope® 5000 Series, Pico Technology, UK). A secondary detection unit, comprising a conversion dynode with an electron multiplier, is located after the Orbitrap electrodes. Its primary function was to facilitate the rough alignment of the ion beam through the apparatus by detecting uncaptured ions.

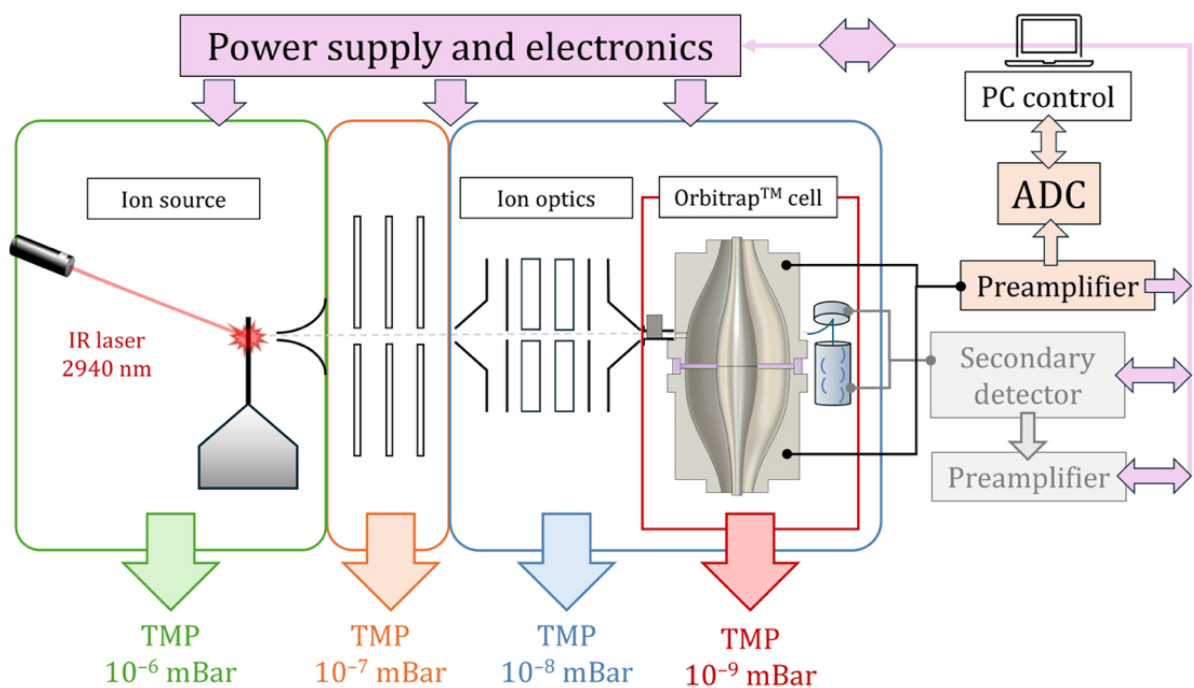

**Figure 1.** Laboratory experimental setup of the HANKA instrument coupled to IR laser ablation chamber.

The absence of a C-trap[33] in our design required a generation of ion packets with well-defined kinetic energy and low angular dispersion for efficient ion trapping and detection within the Orbitrap analyzer. Such conditions were facilitated by a custom-built ion optics based on a previously established concept.[32] The ion optic's principal simulation and assembly is shown in Figure 2. It comprises a series of dedicated electrodes with precise potentials to take care of angular dispersion of ions and to decrease the kinetic energy spread. While the current design provides a good control over the ion kinetic energies, its further optimization may be necessary for the increase of maximum theoretical resolution – which was not the goal here. However, the developed ion optics system offered some advantages over the traditional C-trap modules in terms of complexity. In addition, its relatively robust and compact assembly not only simplifies the overall instrument design but also enables lower power consumption. While the steering ion optics was powered by an in-house designed DC power supply, the switchable electrodes (i.e., guiding electrodes, as well as the central and deflection Orbitrap electrodes) were connected via sequencer and high voltage (HV) switches (Behlke and CGC Instruments, Germany) to HV power supplies (CGC Instruments, Germany; Model 215 Bertan Associates, US).

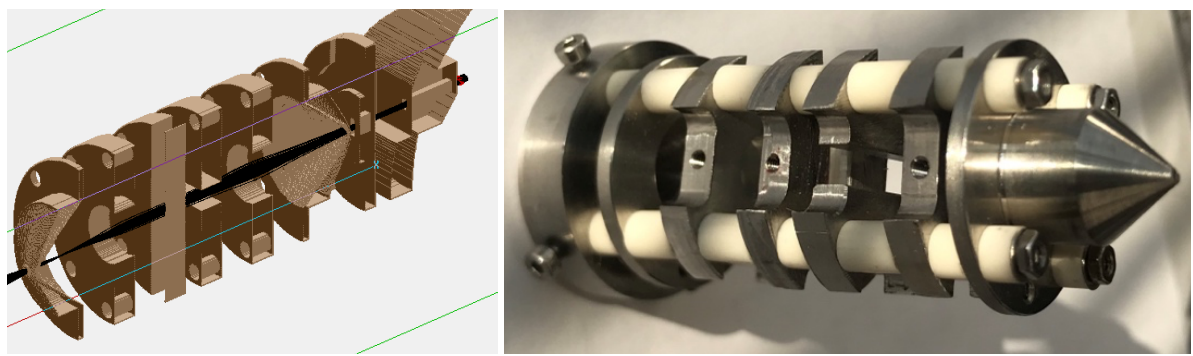

**Figure 2.** Custom build ion optics, SIMION simulation (A), and realization (B).

Two pulse generators (DG 535, Stanford Research Systems, US) were employed to control the timing of the instrument and experiment. The whole system is differentially pumped by four turbomolecular pumps (HiPace, Pfeiffer Vacuum, Germany) resulting in pressures that are indicated in Figure 1.

## 2.2 Experimental setup

To evaluate the performance of the instrument and to optimize parameter settings, a piece of capillary tubing (1/16" OD) made from austenitic chromium-nickel stainless steel grade 304 (1.4301) on a 3-axis manipulator was used as an ablation target. With adjustment of the different parameters including conditions of the laser-induced ablation and plasma formation, all potentials, and timing for ion guiding, focusing, and Orbitrap electrodes, the signal could be optimized in amplitude and resolution. It should be noted that the inner dimensions of cubic modular vacuum chambers correspond to CubeSat dimensions, 10x10x10 cm$^3$. For the small mineral samples, the ionization chamber was equipped with 3-axis manipulator for attaching solid samples. The samples of the meteorites were analyzed using the identical settings as for the stainless-steel tubing. However, the difference between the two sets of measurements was that a piece (Ø 2-3 mm) of the mineral specimen was fixed in a spiral-shape holder from Ag-coated Cu wire, to define its electric potential. The modified setup (with a liquid water beam as a target instead of a solid-state sample in a sample holder) was used to study ice particle impact analogues – here induced by ILAPI. The ionization chamber contained the nozzle on a 3-axis manipulator equipped with a Repeller electrode, used for defining the electric potential of generated ions. A liquid beam with a diameter of approx. Ø 20 µm and 0.5 mL/min flow rate was generated from the nozzle connected to HPLC pump (Knauer GmbH) at a pressure of 50 bar. As a model system for space-relevant organics in water (ice impact analogues) lysine-hydrochloride (100 g/L) in 1 mM hydrogen chloride solution was chosen. The sample at the potential of 1kV was irradiated through a $CaF_2$ lens (f=25 mm) by a pulsed nanosecond (5ns) infrared (2940nm) laser beam (Opolette HE 2731, Opotek Inc., USA) at a repetition rate of 2.2Hz. A variable dielectric attenuator and/or Q-switch delay was used to adjust the output pulse energy to 3mJ (6mJ for liquid phase). The kinetic energy of the produced ions was further increased electrostatically to 1 keV by adjusting the voltages of the sample, as well as the guiding and steering electrodes. Once the ion packet was successfully injected into the Orbitrap analyzer, the periodic motions of ions, i.e., their orbiting around and their axial oscillatory movement along the central electrode, could be detected. The current induced in the outer electrodes was amplified by the differential transimpedance preamplifier and then digitized by a PicoScope. The latter had an integrated deep memory access Field Programmable Gate Array (FPGA) and a board-to-computer communication interface. The pre-processed analog signals were acquired with the sampling rate up to 10 MHz and within time windows of 50 to 500ms.

## 2.3 Data processing

Fast Fourier Transformation (FFT) algorithm was applied to transform the transient signals into frequency domain spectra. The signal was processed for 350ms duration (50ms for analytical purposes, see below), without apodization and zero-filling. Such frequency spectra were converted to mass spectra using the following equation[17]:

$$f = \frac{1}{2\pi}\sqrt{\frac{k}{m/q}} \quad or \quad \frac{m}{q} = \frac{k}{\omega^2} \tag{1}$$

where $\omega = 2\pi f$ is the angular frequency of axial motion, $k$ is the calibration factor (that stems from the curvature of the electrostatic field), and $m/q$ is the mass-to-charge ratio (kg/C). It is then straightforward to express the latter as $m/z$, the mass number-to-charge number ratio according to IUPAC recommendations. Each single shot was calibrated through one-point reference ion self-calibration, where a known (typically most abundant) ion was used as the calibrant for the entire spectrum. The signal processing of the data acquisitions ultimately limited the dynamic range of the single-shot spectrum to 3-4 decades here and to a mass resolution of $R < 70000$ at $m/z = 56$. Mass resolution R (m/Δm at FWHM)**,** mass accuracy (ppm)**,** signal-to-noise ratio (SNR), and the error of the relative isotopic abundance (RIA error) were calculated according to Ref. [24]**.**

**2.4 Materials**

The lunar, Martian, and chondritic materials offered for sale are almost always meteorites (naturally delivered to Earth by impacts), not Apollo or Mars mission returns. Those space agency–retrieved samples remain in scientific custody. Scientists can tell when a meteorite on Earth comes from the Moon or Mars by studying its chemistry, mineralogy, and isotopes, then comparing these traits with samples studied directly by spacecraft. Lunar meteorites, for example, are extremely dehydrated, lack water-bearing minerals, and share the same oxygen isotopic signature as Earth rocks but have unique mineral compositions that match Apollo-returned samples. Martian meteorites are identified most convincingly by tiny bubbles of trapped gas whose composition is identical to Mars's atmosphere as measured by the Viking landers in 1976, along with distinct oxygen isotopes and crystallization ages that differ from asteroidal meteorites. Chondrites, by contrast, are recognized as primitive solar system material because they contain chondrules and isotopic fingerprints that set them apart from Earth, Moon, and Mars. These exotic rocks arrive on Earth naturally: when large asteroids strike the Moon or Mars, the impact energy can accelerate surface fragments above escape velocity, launching them into solar orbit. Over thousands to millions of years, gravitational nudges and sunlight-driven effects alter these orbits until some cross Earth's path, where the planet's gravity captures them. If the fragments survive atmospheric entry, they land as meteorites, often discovered later in deserts or Antarctica. Lunar meteorites tend to reach Earth more quickly because the Moon is nearby, while Martian meteorites may wander through space for millions of years before falling here, yet in both cases, impact ejection provides a natural delivery system that explains how such rare planetary samples end up in collectors' hands. For the experiments, we selected three meteorite samples[34-36]: the Martian shergottite NWA 12269, the lunar melt breccia NWA 11444, and the ordinary chondrite of the L3 group RaS 445. Basic information on these samples is available in the Meteoritical Bulletin Database.[34-36] Current samples were obtained from the above sources and small 1-2mm sized fragments were used without further treatment or purification. To simulate ice grains in space we employed the standard analogue experiment, i.e., we used an IR laser ablation/impact of a micro water beam containing organic molecules, as an analogue to an ice particle impact with the ice particle containing organic material.

## 3. Results and Discussion

The HANKA instrument was developed to apply high-resolution mass spectrometry to laboratory analogue impact experiments[24] – for both, solid-state and ice particles. Its primary goal was to achieve a resolving power, much above R=1000, which is superior to the standard TOF instruments used hitherto. A 10-100 times larger mass resolution, which is realistic for Orbitrap mass sensors over the range of *m/z* = 50 – 250 and beyond, would be a game changer for laboratory analogue experiments and ultimately for future space missions in case of technology matureness. Previous research had focused on analyzing dust impact analogues using time-of-flight instruments similar in spirit (but with larger mass resolution) to spaceborne mass spectrometers like the Cosmic Dust Analyzer (CDA) or the Surface Dust Analyzer (SUDA). The studies demonstrated that laser impact experiments can effectively simulate the impacts of ice dust particles on a target.[13, 24, 37-46] In the present work, the IR laser ablation sampling was applied to a variety of solid-state samples to evaluate the capabilities of high-resolution mass spectrometry, and the simulation of solid particle impacts[47] via laser experiments – now for solid state dust particle <u>and</u> ice particle impacts. The study includes solid-state space samples, i.e., Lunar, Martian, and Chondrite meteorites, and a liquid-beam target with an organic model solute, as a proxy for (organics containing) ice dust particle impacts. A stainless-steel sample served as a defined reference for solid state samples. The performance of the mass analyzer was characterized by the following metrics: mass resolution (m/Δm at FWHM), mass accuracy (ppm), signal-to-noise ratio (SNR), and the error of the relative isotopic abundance (RIA error).

### 3.1 ILAPI of a stainless-steel sample

The single shot mass spectrum of a stainless-steel sample after ILAPI is shown in Figure 3A. The detected mass peaks (isotopes of Cr, Fe, Ni, and Mn) have a mass accuracy around 7 ppm and the mass resolution is close to R > 60,000 (m/Δm at FWHM). When comparing the single-shot spectrum to the averaged spectrum shown in Figure 3B, an improvement in the mass accuracy of the detected peaks (isotopes of Cr, Fe, Ni, Mn, Co, and Cu) is observed. A more detailed analysis of individual spectra and lines showed that the mass accuracy for all identical peaks improves with averaging (Fig. 3B), however, a small variation from shot to shot is noticeable. In consequence, although the accuracy improved by averaging, the overall mass resolution decreased somewhat in the averaged spectrum. In particular, it was observed that the linewidth and thus resolution varies somewhat with individual lines.

Variations in the mass accuracy and resolution of single lines may be attributed to internal Orbitrap cell parameters and small imperfections, including spread of ion kinetic energies, electrode potential ripples, and other deviations from optimal instrument design.[48-50] These factors and also their dynamic fluctuation, may contribute to varying resolutions and uncertainties in the line intensities (for quantification) for different *mass* lines in a spectrum. A minimization of these effects still challenges further instrument development.

A significant improvement of the signal-to-noise level was achieved in the averaged spectrum. This enables the detection of low-intensity peaks associated with the $^{59}$Co$^+$ and $^{63/65}$Cu$^+$. In single shots, mass peaks with a signal-to-noise (SNR) larger than about 1-3 can be detected, whereas in an averaged spectrum mass-lines with much lower intensities can be identified. The dynamic range (DR) of the instrument in its present configuration is approaching $10^4$, which is close but below the theoretical DR of $6.6 \cdot 10^4$. This, in principle, allows us to detect species present at levels well below 1% of the spectrum's main peaks. The theoretical resolving power (for 350 ms signal duration) should be sufficient to distinguish peaks for $^{54}$Cr/$^{54}$Fe (74,000 m/Δm of FWHM), and $^{58}$Fe/$^{58}$Ni (28,000 m/Δm of FWHM). However, in our experiment, only a single peak was observed for each pair. This can be the result of the significant difference in the relative intensities related to the most abundant isotope, namely 2.8 % $^{54}$Cr vs. 6.4% $^{54}$Fe, and 0.3 % $^{58}$Fe vs. 100 % $^{54}$Ni and/or to a small detuning of the instrument.

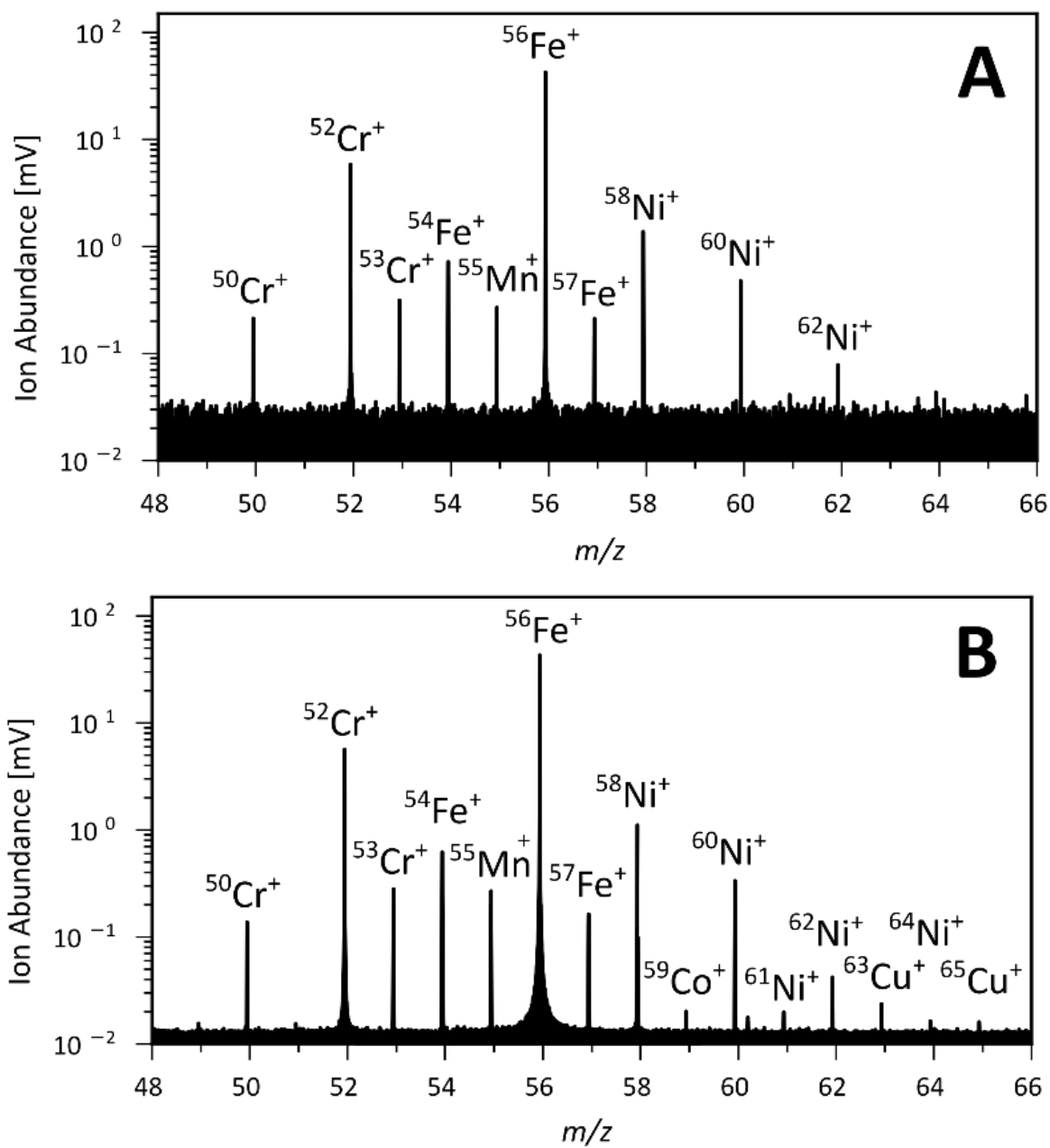

**Figure 3: (A)** Single shot and **(B)** averaged mass spectra with 350 ms signal recording. (A) The base peak $^{56}$Fe$^+$ (97,422 m/Δm at fwhm) was used for mass calibration. All detected spectral lines were assigned to atomic elements with an accuracy 7 ppm, with a mass resolution well above 60,000 m/Δm at m/z range 50-65. (B) In the averaged spectrum with a significantly decreased noise (64 spectra) the base peak $^{56}$Fe$^+$(88,912 m/Δm at fwhm) was used for mass calibration. All detected spectral lines were assigned to atomic elements with an accuracy of <6 ppm. The mass resolution was determined to be <30,000 m/Δm at m/z range 50-65. The displayed mass spectra are log-scaled and without baseline correction.

Whether the low-abundant species lines ($^{54}$Cr and $^{58}$Fe) are masked though overlap with more intense lines or hidden in the noise through local variations in the linewidths is unclear at present. For a dust particle impact event in space, it is important that a mass analyzer can resolve a spectrum from this single event, i.e., from a single shot. Based on the above results, the sensitivity of our instrument, that is planned to be a dust particle mass analyzer, appears to be sufficient in this context. In particular, the use of the HANKA instrument in laboratory impact experiments (employing LILBID or SELINA technology) appears feasible. The averaging of mass spectra may have an advantage in exploratory laboratory analogue experiments because the overall dynamic range and SNR is improved. However, in the single shot/impact spectra several sources of errors may appear, i.e., laser intensity and ILAPI variations, as well as sample inhomogeneities upon ablation, and finally imperfections of the Orbitrap alignment. Therefore, the intensities in the mass spectra are expected to vary to some extent "naturally".

The RIA error for the single shot spectra was in reasonable agreement with the expectations. It was determined to be within 10-30% (in a single short/impact experiment), calculated from the experimentally determined and the theoretical relative isotope abundances. Very low abundant isotopes (RIA <15 %) in the presence of high abundance isotopes, may even display higher errors as much as 30%. The determination of accurate relative isotopic abundances with an Orbitrap instrument is a known challenge. Relative isotopic abundances measured with Orbitrap instruments can be biased when isotopologue abundances differ greatly. Space charge effects, differential coherence decay, and limited detector dynamic range can cause underestimation of low-abundance isotopes, while effects are minimal when isotopes are of similar abundance. These distortions likely arise from a combination of ion–ion interactions and instrumental limitations. Whether the observed error (in our case here) is due to instrument imperfections or otherwise unwanted ion interaction in the orbitrap, is the topic of ongoing investigations in our lab. The correlation between shorter ion trajectories and low abundance discrimination has been qualitatively seen, when we evaluate smaller time-windows (50ms only as opposed to 350ms) of the transients and process shorter time windows in the Fourier transform (FT) data processing. For the smaller time windows the dynamic range and sensitivity were poorer, but the isotope ratios were closer to the theoretical expectations, supporting the above conclusions. We will come back to this specific issue in chapter 3.4 below. In conclusion of this chapter, we find that the present Orbitrap mass sensor has a good single shot/event sensitivity and a very good mass accuracy and resolution being the key parameters for a dust detector in space. The uncertainties in the line intensities in single shot/impact events are due to the variations of the process and some imperfections in the Orbitrap tuning, which are found on the same order of magnitude.

### 3.2 ILAPI of meteorite minerals

Mass spectra obtained from three distinct meteorite samples, demonstrate the high sensitivity and overall performance of our prototype instrument showing the broad analytical capabilities for dust impact material analysis.[47] The overall appearance of elementary cations in the spectra indicate that the ILAPI method employed here may also be a suitable analogue experiment for hypervelocity dust impacts with cold plasma ionization, with velocities > 5 km/s. Like in the case of ice grain impacts the ILAPI may be an equally suitable technique for solid dust impacts. ILAPI mass spectra can directly be compared with hypervelocity impacts measured in the large dust accelerators at the University of Boulder Colorado,[54] or at the University of Stuttgart.[2] A prerequisite, however, is a calibration which correlates laser energies with particle velocities – as has been demonstrated for ice particles before.[40] The universal employment of an IR laser for the ILAPI and its dual use for solid-state and ice particles makes our approach unique. Note, employing an IR laser in space applications is not necessary anymore. We expect that this methodology will be directly relevant for future dust analyzer instruments on flyby missions, providing a laboratory analog for in-situ analysis.

Single shot mass spectra of Chondritic, Martian and Lunar meteorites are shown in Figures 4-6. The spectra all displayed a mass resolution R ≥ 60,000 within the *m/z* range 12-60. The instrument's dynamic range reaching 3-4 orders of magnitude, allows the simultaneous detection of highly abundant to nearly trace elements. The detected ions stem mostly from inorganic-mineral formatting elements, including their stable isotopes, such as Na, Mg, Al, Si, K, Ca, and Fe.

The Lunar regolith (Figure 4) is classified as melt breccia. Its mineral composition is dominated by silicate- and oxide- forming elements (e.g. Na, Mg, Al, Ca, Fe, Ni), with distinct signatures reflecting basaltic lunar regolith composition. All these elements were present and detected in single shot spectra at a mass precision of 7.2 ppm. Besides mineral-forming elements, other elements such as carbon, oxygen, and sulfur were also confirmed. Thanks to the instrument dynamic range, nearly all stable isotopes of calcium (up to $^{43}$Ca), except the least abundant $^{46}$Ca could be detected.

The Martian shergottite sample (Figure 5) shows mineral composition like the lunar meteorite. This sample exhibits notable enrichment in Fe- and Mn- bearing minerals, which are indicative of formation under oxidizing and volcanic conditions, and volcanic or subsequent hydrothermal alteration. Presented findings are consistent with observations from other Martian meteorites and rover analyses.[34-36] The chemical composition includes common mineral-forming elements along with phosphorus. Detected spectral lines were identified with a mass accuracy of 5.1 ppm.

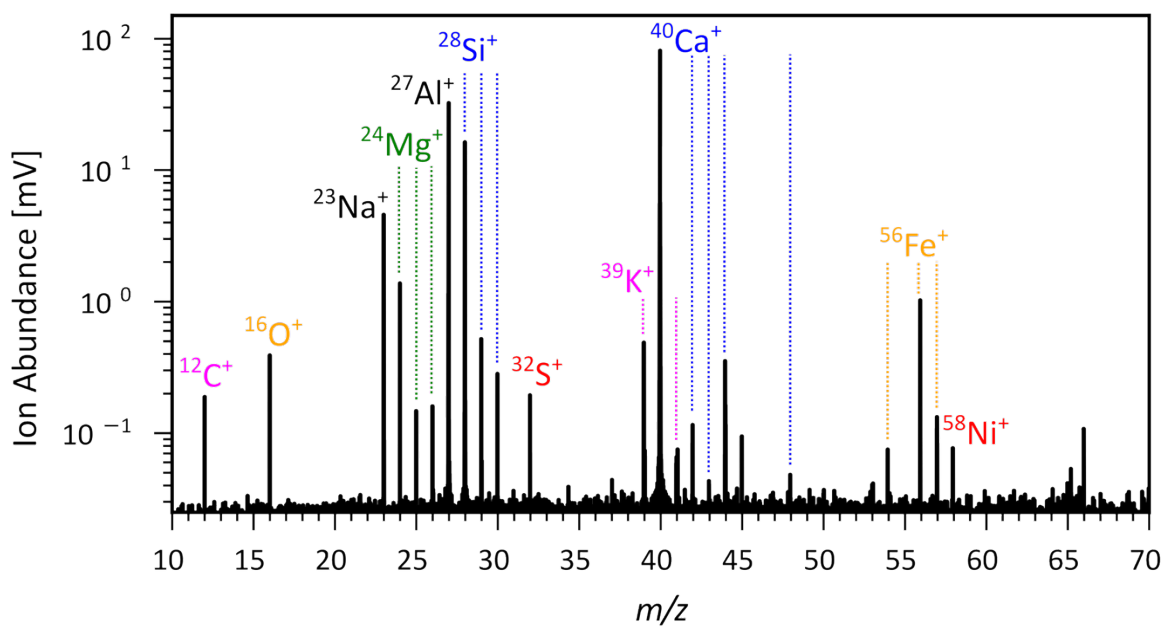

**Figure 4.** Single shot mass spectrum of Lunar meteorite (NWA 11444) with 350 ms signal duration. The presented mass spectrum is log-scaled and without baseline correction. The base peak $^{40}$Ca$^+$ (R=81,435) was used for mass calibration. All detected spectral lines were assigned to atomic elements with an accuracy of 7 ppm, and with a mass resolution R>60,000 in the mass range 12-65. The most abundant isotope of each element is labeled with the color-coded dotted lines corresponding to other isotopologues.

The mass spectrum of the Martian sample contains four unassigned spectral lines. We attempt a preliminary assignment by taking advantage of the high resolution and line precision. In particular, the line at *m/z* 46.9679, corresponding to a mono cation agrees well with an assignment to OP$^+$. Remaining unassigned peaks were limited to only 3 to 5 suggestions of elemental compositions. The line at *m/z* 44.9788 may correspond to elemental compositions such as SiOH$^+$ and CHS$^+$. The line, at *m/z* 56.9657, possible assignments may be CaOH$^+$, and HSMg$^+$. The latter two peaks were also detected in the mass spectrum of lunar meteorite sample. Nevertheless, the line at *m/z* 22.4208 resisted identification, since the mass was not consistent with any of the expected and considered cations.

The Chondrite sample (Figure 6) belongs to ordinary chondrite class L3. The high content of Fe and Ni metal ions is obvious at first sight of the spectrum. The overall mineral composition is consisted with common silicate- and oxide- forming elements. In addition, carbon, oxygen, and sulfur was also detected. In the mass spectrum of Fig. 6, an unassigned line at *m/z* 52.9845, mostly agrees with an elemental composition of C$_2$HSi$^+$ and/or AlCN$^+$. Mass lines at *m/z* = 35.2961 and 13.7497 could not be assigned yet, since it did not match any *expected* cation mass within our mass accuracy range.

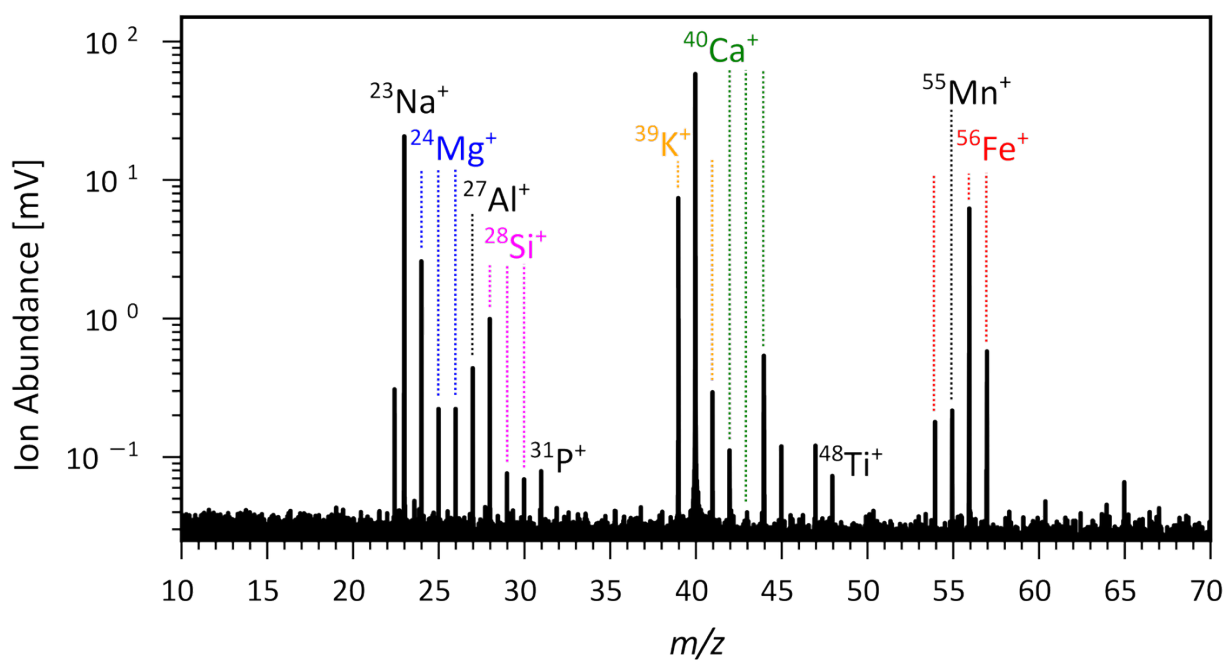

**Figure 5.** Single shot mass spectrum of Martian meteorite (NWA 12269) with 350ms signal duration. The presented mass spectrum is log-scaled and without baseline correction. The base peak $^{40}$Ca$^+$ (R=53,388) was used for mass calibration. All detected spectral lines were assigned to atomic elements with an accuracy of 7 ppm, with mass resolution R>60,000 in the mass range 12-65. The most abundant isotope of each element is labeled with the color-coded dotted lines corresponding to other isotopologues.

It should be noted that the determined relative isotopic intensities of the 3 mineral samples are in fair agreement (10-30%) with expected values, even in this single shot configuration (within the shot-to-shot variations of intensities of the experiment). The (systematic) deviations from theoretical isotope ratios we attribute to systematic biases when isotopologues differ markedly in abundance. High-abundance species can dominate the Coulombic field within the analyzer, perturbing the trajectories and oscillation frequencies of co-trapped low-abundance ions (space charge effects), which leads to frequency shifts and signal attenuation of minor isotopologues. Differences in coherence decay rates can further amplify this, i.e., low-abundance ions often lose phase coherence more rapidly, reducing their detectable image current relative to dominant species.

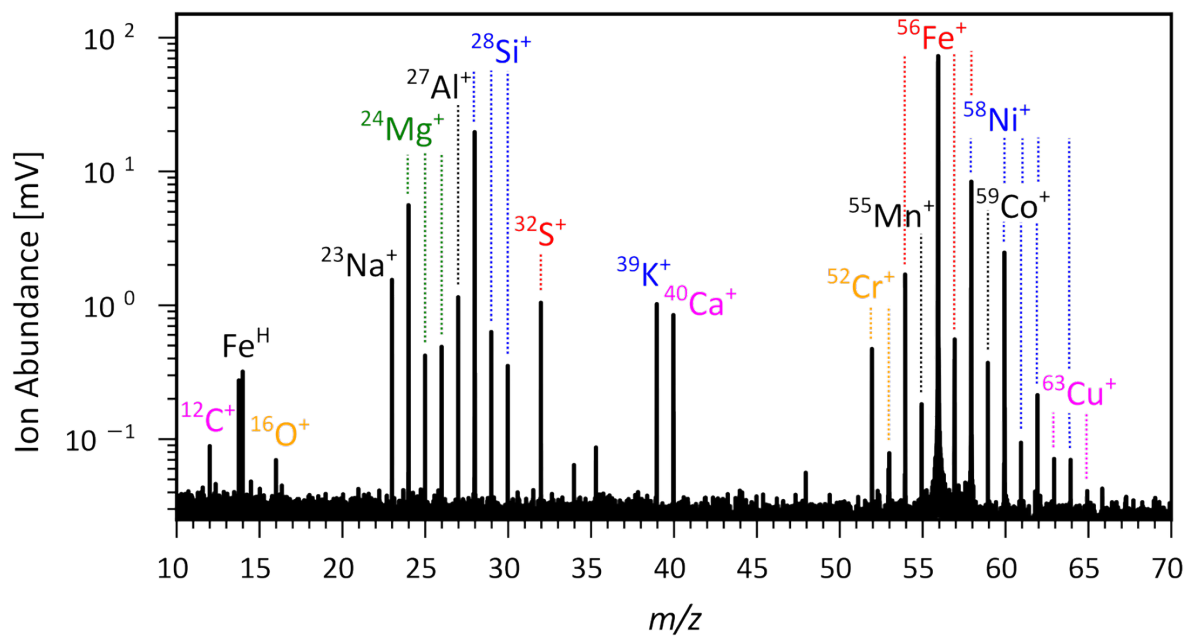

**Figure 6.** Single shot mass spectrum of Chondrite meteorite (RAS 445) with 350 ms signal duration. The presented mass spectrum is log-scaled and without baseline correction. The base peak $^{56}Fe^+$ (R=73,386) was used for mass calibration. All detected spectral lines were assigned to atomic elements with accuracy 11 ppm, with a mass resolution R~60,000 in the mass range 12-65. Note: $Fe^H$ is an artefact and attributed to harmonics. The most abundant isotope of each element is labeled with the color-coded dotted lines corresponding to other isotopologues.

Our comparative analysis of the above materials displays the key geochemical differences among the Moon, Mars, and primitive meteorite bodies. Variations in element composition, their oxidation states, volatile content, and element partitioning reflect the differing thermal and geological histories of their parent bodies. For example, iron speciation in Martian analogs points to special redox conditions, whereas the reduced nature of lunar samples may support long-term exposure to a vacuum environment. The ability to resolve these differences from minimal material, using a contactless ablation approach, demonstrates the power of HANKA for dust impact analysis.

Although the current study focused primarily on molecular and elemental identification, the observed resolution (R≥60,000), the single-impact sensitivity, and the achieved mass accuracy suggest future application possibilities, e.g., in laboratory dust impact experiments, and in the future, even as a miniature dust detector in space. In particular, HANKA's resolution and mass accuracy may enable isotopic ratio measurements in its advanced version in the future. This may be critical and a game changer for planetary dating, and for tracing nucleosynthetic anomalies. This may position HANKA not only as a tool for chemical composition characterization of dust impacts but also for chronometry and space exploration studies *in situ* in the future.

**3.3 ILAPI on a liquid beam in vacuum as a proxy of ice dust impact**

As outlined above, several previous studies have shown, that IR laser desorption and ILAPI is a good laboratory analogue for ice particle impacts and therefore for the analysis of impact mass spectra in space such as from missions like Cassini and Europa Clipper.[37-46] The previous results underscore the importance of high-resolution mass spectrometry in laboratory analogue mass spectrometry. This is in particularly true for the detection of biosignatures in future missions.

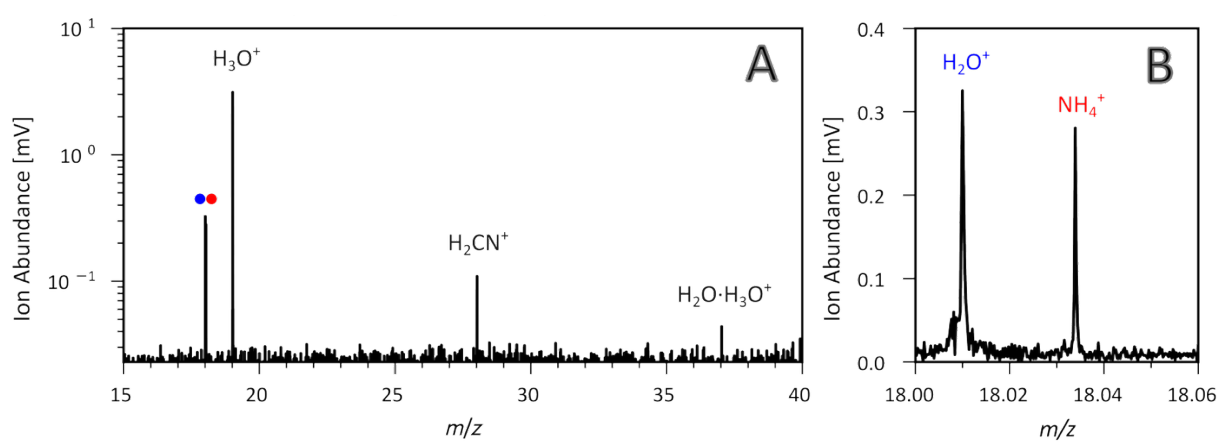

**Figure 7.** Single shot mass spectrum with 100ms signal duration of ILAPI of an amino acid (L-Lysine) containing water filament (100 g/L in 1 mM hydrogen chloride solution) in vacuum, simulating ice particle impact in the laboratory. A) overview of the single shot mass spectrum of the L-Lysine solution; B) details of the $m/z$ = 18 region.

Here, a high-resolution Orbitrap-type instrument such as HANKA would be advantageous. Recently, we have shown that an Orbitrap-type instrument can be combined with liquid beam desorption – that we call ILAPI here.[24] In the following we show first experiments in which we combined HANKA and its ion optics with our LILBID experiment. The ILAPI on a liquid µ-beam of water containing L-lysine as a model organic compound shows again the high-resolution performance of the system and superior dynamic range, with respect to previous results.[24] We detected water clusters and fragments originating from IR-laser impact on the L-lysine containing water filament mimicking the ice particle impact. As confirmed earlier, amino acids display characteristic energy dependent fragmentation patterns. The resolution of assigned peaks within the present single shot mass spectrum was R≥35,000 at *m/z* 18-37. The average mass accuracy was determined to be 7.5 ppm.

### 3.4 Determination of Relative Isotopic Abundances with Orbitrap Mass Analyzers

Here we come back to the accurate quantification of isotopes in an Orbitrap instrument. Accurate determination of relative isotopic abundances using Orbitrap mass analyzers remains a recognized analytical challenge.[48-50] While these instruments offer exceptional mass resolution and accuracy, several inherent physical and instrumental factors can distort measured isotope ratios, particularly when isotopologues differ greatly in abundance. The most significant contributor is the influence of space charge.[51] When a high number of ions are injected simultaneously, their mutual Coulombic repulsion can perturb the electrostatic field inside the analyzer. Abundant isotopologues dominate this field, which can alter the trajectories and oscillation frequencies of co-trapped, low-abundance ions. This phenomenon can lead to frequency shifts and reduced apparent intensities of the minor isotopologues, thereby biasing relative abundance measurements.

In addition to space charge, differences in ion coherence and decay dynamics play a role.[48-50] The Orbitrap detects ions based on the coherent image current they induce, and even a small coherence decay of the group motion of ions can result in slightly different parameters of the image current oscillation function. Minor isotopologues, often fewer in number and more susceptible to trajectory perturbations, may lose coherence more rapidly. This faster decay diminishes their signal relative to the more stable and intense signal from the major isotopologues. Injection and trapping processes can also contribute - small variations in ion energy, spatial distribution, or injection timing can lead to preferential trapping and longer survival of the more abundant species. Furthermore, limitations in the dynamic range of the detection electronics and the Fourier transform processing can cause the strong signals from abundant isotopologues to overshadow weaker signals, especially as the transient progresses and the signal-to-noise ratio declines.

These effects are typically less pronounced when isotope ions are present in similar abundances, as they exert comparable influence on the space charge field and coherence decay behavior, reducing relative bias. However, when abundance differences are large, the dominant isotope effectively governs the local ion environment, resulting in systematic underestimation of the minor isotope species. Whether these distortions arise primarily from instrumental imperfections or from unavoidable ion–ion interactions within the analyzer are still under investigation. Nevertheless, it must be kept in mind that the shot-to-shot variations in the intensities of the mass lines for single impacts are another source of the line intensity uncertainties of the present method (and in real applications, such as single particle impacts in the laboratory and in space), therefore even with perfect Orbitrap tuning some uncertainties will remain. For an overall minimization of these effects current research efforts are directed toward optimizing ion populations, optimizing transient acquisition parameters with online control of Orbitrap parameters and potnetials, and post-acquisition corrections to mitigate these effects.

### 3.5 Towards a Universal Dust Detector

From a technical perspective, the present results validate the integration of IR laser ablation and thus ILAPI with Orbitrap mass spectrometry in laboratory analogue experiments for robust, high-resolution analysis of solid and ice particle dust impacts (analogues of "space dust"). Employing a single IR laser for ILAPI enabled for the first time a universal process (ILAPI) for the simulation of impacts of very different sorts of space dust, i.e., mineral dust and ice particles.[52, 53, 54] The analogue impact process could be analyzed with HANKA, an compact Orbitrap-type mass analyzer.

Although, ILAPI was used for the first time here to mimic solid and ice dust impacts in the laboratory the IR laser is employed only for an optimal analogue experiment simulating and understanding the impact event in the laboratory - it is not necessary in space on a spacecraft detecting space dust. This is a unique selling point of our approach. The HANKA Orbitrap detector including target, preamp and ion optics (attached to the outside wall) is made to fit into a standard ideal vacuum cube. The power supply and further electronics is in preparation for another IdealVacuum$^{TM}$ cube. The target size will two CubeSat units in the near future. Miniaturized space-ready switching electronics are being developed at the moment. In summary, we find that the HANKA technology is at an intermediate technology readiness level (TRL4). At TRL 4, individual components and subsystems of HANKA are combined in CubeSat-type cubes and extensively tested in a controlled laboratory environment to validate performance at present, which is a prototype and thus a simplified version of a final flight instrument. A special geometry impact target at present undergoes tests with hypervelocity solid particles at the University of Boulder Colorado.[54]

The main challenge/argument of flying an Orbitrap mass spectrometer in space is not its scientific value but its engineering maturity. While Orbitraps are well established in laboratories, their technology readiness level for space remains lower because they have only recently been miniaturized and adapted for operation in vacuum, radiation, and long-duration missions.[19,20] While traditional laboratory versions are large and power-hungry, spacecraft designs require compact, efficient variants that still preserve high resolution. In addition, Orbitraps rely on stable voltages, precise ion optics, and vibration-free conditions, all of which are harder to guarantee in the harsh thermal and mechanical environment of space. Space missions also tend to prefer instruments with strong flight heritage, such as quadrupole or time-of-flight spectrometers, making the Orbitrap analyzer appear higher risk despite its superior capabilities. In short, the barriers are mainly the lower TRL, resource constraints, and lack of flight history, but as prototypes prove themselves in relevant environments, Orbitraps are expected to become standard tools for planetary exploration.

The compact design and passive sampling approach—requiring no sample manipulation or preparation—may make HANKA a strong high resolution mass detector candidate for upcoming missions such as the L4 ESA mission back to Enceladus after 2040.[55] Whether aboard lunar orbiters, asteroid landers, or comet flybys, the system may offer a high performance low-resource path to rich molecular data. The Orbitrap mass spectrometer offers capabilities that make it uniquely valuable for planetary exploration. Its greatest strength is high mass resolution and accuracy, allowing scientists to distinguish molecules with nearly identical masses and to assign exact molecular formulas. This precision is essential for detecting complex organics and potential biosignatures in environments like Europa's plumes or Enceladus's geysers.

Unlike simpler instruments, the Orbitrap can analyze a wide range of samples—gases, ices, or solids—and is sensitive enough to detect trace compounds even in low-density environments. Its versatility makes it ideal for missions that must rely on brief encounters, such as flybys through icy moon plumes. Adapted designs are now robust for spaceflight, with miniaturization, power efficiency, and durability in vacuum and radiation-rich settings. These traits support long missions far from Earth while maintaining data quality. Equally important, as has been shown here, Orbitrap data can be compared directly with terrestrial laboratory results, creating strong continuity between Earth-based research and in situ exploration. This consistency helps ensure that novel molecules or isotopic signatures discovered in space can be confidently interpreted. In short, the Orbitrap provides the sensitivity, versatility, and reliability needed to transform planetary chemistry studies, advancing the search for habitability and life beyond Earth.

## 4. Conclusions and Outlook

In this study, we demonstrated the application of a compact Orbitrap-based mass spectrometer—HANKA (*High-resolution mass Analyzer for Nano-scale Kinetic Astro materials*)—for high-resolution and sensitive ("single event/particle") chemical analysis of planetary analog materials using infrared laser ablation and plasma ionization. Mass spectra obtained from Lunar, Martian, and Meteoritic samples consistently exhibited a resolution of approximately 60,000 (FWHM), enabling high-precision identification of elements and molecular species. These results provide strong experimental support for using Orbitrap technology as a powerful method for in situ characterization of cosmic dust.

HANKA represents a significant advancement over previous Earth-based laboratory analogue experiments and displays a high potential for future applications in space as a universal dust detector for the compositional analysis of solid state and ice containing dust particles. Its compact dimensions, high sensitivity, the superior mass resolution and -accuracy, and its ability to analyze solid materials without extensive sample preparation positions itself as a promising candidate for future missions targeting dusty environments—such as the surfaces and exospheres of airless bodies, cometary comae, asteroid regolith, and interstellar dust streams.

Looking ahead, HANKA has the potential to serve as a universal dust detector for space applications. Its passive, impact-based or laser-driven sampling strategy is adaptable to a wide range of mission profiles, from low-gravity surface operations to high-velocity flybys. By delivering molecular-level insight into the composition and origin of dust particles, HANKA can help address fundamental questions about planetary evolution,[56] prebiotic chemistry, and the transport of organic materials across the solar system. The present concept appears to be well suited for the laboratory analysis of complex chemical compositions and biosignatures in ice grains of the Europa Clipper mission,[57] or (with a high TRL) even as an advanced dust detector concept in future missions such as ESA's L4 Enceladus Orbilander mission planned for 2040.[55]

The roadmap for future work includes environmental testing under space-relevant conditions, comprehensive testing with hypervelocity ice and mineral particles with SELINA[52,53] and the Boulder Dust Accelerator.[54] As outlined above, relative isotopic abundances measured with Orbitrap instruments can be biased when isotopologue abundances differ greatly.[58] Space charge effects, differential coherence decay, and limited detector dynamic range can cause underestimation of low-abundance isotopes, while effects are minimal when isotopes are of similar abundance. These distortions likely arise from a combination of ion–ion interactions and instrumental limitations and should be resolved when Orbitrap-based mass sensors are ready for missions to interpret isotopic data. Complete design incorporating impact target and space-ready electronics in HANKA can be then integrated into flight-qualified platforms, and mission concept development for lunar, Martian, and small-body exploration. As we move toward more ambitious missions that explore the chemical diversity of our solar system and beyond, instruments like HANKA will be essential in transforming each microscopic dust grain into a unique and rich data window on the cosmic objects.

As a final note we want to emphasize that the observed small imperfections of the present setup (intensity variations for isotopes and close masses with large differences in abundance) likely stem from small (transport) detuning of the home build orbitrap instrument. This is the reason why the current instrument does not yet reach specifications like those of commercial instruments. However, in single particle/shot impacts also variations of the ideal bulk concentrations of the samples are expected. They differ even for different impact conditions from particle to particle. This variation may be as large as the somewhat limited instrument precision here. In conclusion, we highlight here, and we point out that the current study was to show that the instrument is robust and its sensitivity in single impact events is high enough, as well as the instrument resolution can be well above R=50000, which would be sufficient for future space applications. Whether the optimization and tuning of an orbitrap on earth can be maintained in space after a rocket start with vibrations and non-standard conditions (temperatures) remains to be seen in future investigations. This will then be the topic of instrument developments toward higher technology readiness levels in our laboratory (TRLs).

**Acknowledgements**

We thank Prof. Dr. A. Makarov (*Thermo*, Bremen) for fruitful discussions and helpful comments within the project and on the manuscript. The authors acknowledge financial support from the German Science Foundation (DFG) through grant AB 63/25-1, and of the Czech Science Foundation (GAČR) through grant No. 24-13757L. BA thanks the European Union for funding within the ERA-Chair Project 101186661 SPACE. MF would like to acknowledge the financial support of the Technology Agency of the Czech Republic (TAČR) under the grant of the National Centre of Competence for Aeronautics and Space, reg. No. TN02000009/11 FREYA II.

**Contributions of Authors**

Conceptualization: BA, JZ, IZ, AS, MP, BC, ML
software: IZ, MM, AS
validation: BA, AC
formal analysis: IZ, MM
resources: BA, JZ, MF
data curation: MM, IZ, AC
writing: BA, MM
writing–review and editing: all
visualization: MM
supervision: BA, JZ, IZ
project administration: BA
funding acquisition: BA, JZ

All authors (M. Malečková, I. Zymak, A. Spesyvyi, M. Polášek, B. Cherville, M. Lacko, M. Ferus, A. Charvat, J. Žabka, and B. Abel) have read and agreed to the published version of the manuscript.